# Few-photon computed x-ray imaging

## Zheyuan Zhu,[1,*] Shuo Pang[1]

[1]*CREOL, The College of Optics and Photonics, University of Central Florida, 4304 Scorpius St, Orlando, FL, 32816, USA*
*\* zyzhu@knights.ucf.edu*

**Abstract:** X-ray is a ubiquitous imaging modality in clinical diagnostics and industrial inspections, thanks to its high penetration power. Conventional x-ray imaging system, equipped with energy-integrating detectors, collects approximately $10^3 - 10^4$ counts per pixel to ensure sufficient signal to noise ratio (SNR). The recent development of energy sensitive photon counting detectors opens new possibilities for x-ray imaging at low photon flux. In this paper, we report a novel photon-counting scheme that records the time stamp of individual photons, which follows a negative binomial distribution, and demonstrate the reconstruction based on the few-photon statistics. The projection and tomography reconstruction from measurements of ~ 16 photons show the potential of using photon counting detectors for dose-efficient x-ray imaging systems.

## 1. Introduction

Due to its high penetrating power, x-ray imaging is extensively used as a non-invasive imaging method in medical diagnosis and industrial inspections. Among many x-ray contrast mechanisms, such as phase contrast [1] and coherent scattering [2] etc., x-ray imaging modalities based on attenuation, such as radiography (projection) and computed tomography (CT), remain the most common ones. Conventional x-ray imaging systems count $10^3$-$10^4$ photons per pixel in a fixed period. To ensure the signal to noise ratio (SNR), high radiation dose is administrated to the sample [3], prohibiting the imaging of objects that are susceptible to radiation damage. Conformational change due to high radiation dose is a major concern for imaging of biological samples [4,5]. Microprocessors and flash memories are also vulnerable to physical damages under excessive x-ray radiation [6,7]. In the field of both biomedical diagnosis and industrial inspection, developing an x-ray imaging system at extremely low photon flux without sacrificing its quality is highly desirable [8,9].

In visible and infrared optical imaging regime, the use of avalanche photodiodes to time-resolve the single-photon events allows the range and reflectivity imaging at a few photons per pixel [10,11]. X-ray detector with single-photon sensitivity has opened new opportunities for photon-efficient imaging in medical CT and integrated circuit inspections [12,13], yet its current usage is limited to the traditional time-integration mode, which counts the total number of photons in a predefined integration time. Here we report a novel photon-counting scheme that records the time stamp of individual x-ray photons, which follows a negative binomial distribution. We have demonstrated the reconstruction under low photon flux by taking the few-photon statistics into consideration.

## 2. Theory

In few-photon detection regime, instead of collecting the total number of photon per pixel, our method records the number of time intervals elapsed between two adjacent photon events. Fig. 1 shows the concept of operation. The time stamps of individual photons collected from the Si-PIN detector are registered by a data-acquisition device. Let $\lambda$ be the probability of receiving one photon in each time interval, $\Delta t$, when no sample was present. Considering the sample attenuation, for each pencil beam, $j$, the probability of receiving one photon in each time interval is

$$T_j = \lambda \exp\left(-\sum_{i=1}^{n} \mathbf{A}_{ij}\mathbf{f}_i\right) \tag{1}$$

where the subscript $i = 1,2,\ldots,n$ represents the index of the discretized object attenuation map $\mathbf{f}$; and $j = 1,2\ldots m$ represents the index of discretized pencil-beam measurements. The matrix $\mathbf{A}$ establishes the linear relation between the object and the measurement. For x-ray projection imaging, $\mathbf{A}$ is the unitary matrix, and for tomography, $\mathbf{A}$ represents the Radon transform matrix constructed from the distance-driven ray-tracing model [14].

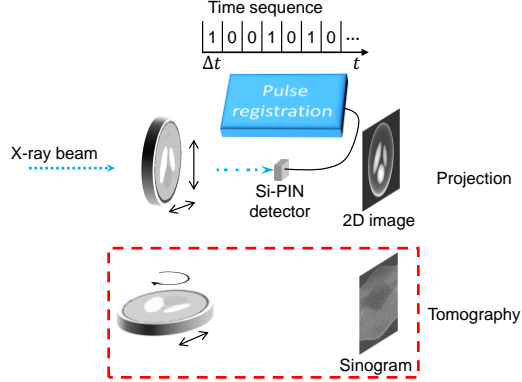

Fig. 1: Illustration of the photon-counting scheme for both x-ray projection and tomography imaging.

The joint probability of detecting the $\mathbf{r}$-th photon at $\mathbf{g}$-th time imterval follows the negative binomial distribution $\mathbf{g} \sim NB(\mathbf{r}, \mathbf{T})$, whose probability mass function (PMF) is

$$p(\mathbf{g}|\mathbf{f};\mathbf{r}) = \prod_{j=1}^{m} \binom{g_j-1}{r_j-1}\left(1 - \lambda\exp\left(-\sum_{i=1}^{n}\mathbf{A}_{ij}\mathbf{f}_i\right)\right)^{g_j-r_j}\left[\lambda\exp\left(-\sum_{i=1}^{n}\mathbf{A}_{ij}\mathbf{f}_i\right)\right]^{r_j} \tag{2}$$

where $\mathbf{g} = (g_1, g_2, \ldots g_m)$ is the total number of time intervals that has elapsed upon the arrival of $\mathbf{r} = (r_1, r_2, \ldots r_m)$ photons at each pencil beam, $j$. In contrast, conventional photon-counting scheme records the number of photons in a predefined period $\mathbf{g}\Delta t$, during which the joint probability of receiving $\mathbf{r}$ phtons for each penil beam is binomially distributed $\mathbf{r} \sim B(\mathbf{g}, \mathbf{T})$

$$p(\mathbf{r}|\mathbf{f};\mathbf{g}) = \prod_{j=1}^{m} \binom{g_j}{r_j}\left(1 - \lambda\exp\left(-\sum_{i=1}^{n}\mathbf{A}_{ij}\mathbf{f}_i\right)\right)^{g_j-r_j}\left[\lambda\exp\left(-\sum_{i=1}^{n}\mathbf{A}_{ij}\mathbf{f}_i\right)\right]^{r_j} \tag{3}$$

The reconstruction is an optimization problem that minimizes the negative log-posterior distribution

$$\hat{\mathbf{f}}(\mathbf{g};\mathbf{r}) = \underset{\mathbf{f}}{\mathrm{argmin}}\{l(\mathbf{f}) + \tau TV(\mathbf{f})\} \tag{4}$$

The objective function consists of two parts: the first part is the negative log-likelihood of the distribution $l(\mathbf{f}) = -\log p(\mathbf{g}|\mathbf{f};\mathbf{r})$; and the second part $TV(\mathbf{f})$ is a total-variance (TV) regularizer [15] with a non-negative parameter $\tau$. After neglecting constant terms independent on $\mathbf{f}$, both conventional and our time-stamp photon counting schemes have the same negative log-likelihood

$$l(\mathbf{f}) = \sum_{j=1}^{m}\left\{r_j \sum_{i=1}^{n}\mathbf{A}_{ij}\mathbf{f}_i - (g_j - r_j)\log\left[1 - \lambda\exp\left(-\sum_{i=1}^{n}\mathbf{A}_{ij}\mathbf{f}_i\right)\right]\right\} \quad (5)$$

with gradient

$$\nabla l(\mathbf{f}) = \mathbf{A}^T\left(\mathbf{r} - \frac{\lambda(\mathbf{g}-\mathbf{r})\exp(-\mathbf{Af})}{1-\lambda\exp(-\mathbf{Af})}\right) \quad (6)$$

and Hessian matrix

$$\mathbf{H}(l(\mathbf{f})) = \mathbf{A}^T\left(\frac{\lambda(\mathbf{g}-\mathbf{r})\exp(-\mathbf{Af})}{(1-\lambda\exp(-\mathbf{Af}))^2}\right)\mathbf{A} \quad (7)$$

The noise of the most widely-used panel detector, which consists of a scintillator optically coupled to a CMOS or CCD camera, is dominated by a Gaussian distribution $N(\mu_j, \sigma)$ with mean

$$\mu_j = I_0 \exp\left(-\sum_{i=1}^{n}\mathbf{A}_{ij}\mathbf{f}_i\right) \quad (8)$$

where $I_0$ represents the detector readout of the un-attenuated X-ray beam. The variance $\sigma$ is assumed uniform for all pixels and is calibrated with 30 snapshots from the detector. The negative log-likelihood of observing the readout of each measurement, **y**, given the object **f** is

$$l(\mathbf{f}) = \sum_{j=1}^{m}\left[y_j - I_0 \exp\left(-\sum_{i=1}^{n}\mathbf{A}_{ij}\mathbf{f}_i\right)\right]^2 /(2\sigma^2) \quad (9)$$

with gradient

$$\nabla l(\mathbf{f}) = \mathbf{A}^T\left\{\frac{I_0}{\sigma^2}[\mathbf{y} - I_0\exp(-\mathbf{Af})]\exp(-\mathbf{Af})\right\} \quad (10)$$

and Hessian matrix

$$\mathbf{H}(l(\mathbf{f})) = \mathbf{A}^T\left(\frac{I_0}{\sigma^2}\exp(-\mathbf{Af})(2I_0\exp(-\mathbf{Af}) - \mathbf{y})\right)\mathbf{A} \quad (11)$$

All the operations except $\mathbf{A}(\cdot)$ or $\mathbf{A}^T(\cdot)$ in Eq. (6-7, 10-11) shall be interpreted as element-wise.

## 3. Material and methods

### 3.1 Experiment setup

The photon-counting projection (PC-projection) and tomography (PC-CT) imaging used a copper-anode x-ray source (XRT60, Proto Manufacturing) operating at 12kV, 1mA. This low power setting avoided the overlap of two photon incidences within one time interval when no sample is present. The x-ray beam was collimated by a pair of 0.5mm pinholes to form a pencil-beam illumination. A Si-PIN detector (X-123, AMPTEK) was connected to a data-acquisition (DAQ) device (USB-6353, National Instrument) programmed in the edge-counting mode. The low-energy channels (<1keV) on the detector were filtered out to eliminate the dark noise. The output of the DAQ device was a series of $\Delta t$=10μs time intervals, within which either one photon or zero photon was registered. In PC-projection, the sample was translated both horizontally and vertically across the beam by two linear stages (UTM150CC, Newport). In PC-CT the sample was also rotated 180° around the vertical axis by a rotational stage (RV1200P, Newport).

Before CT scans, the noise model of the photon-counting system and the incident photon flux $\lambda$ were calibrated with a projection measurement on a linear attenuation pattern, which was created by stacking multiple paper layers with identical thickness $h = 0.12\text{mm}$. The pattern was divided in to 9 (3 X 3) regions, with region 1 being air and region 9 corresponding to 8 paper layers. For CT scan, a laser-machined acrylic resolution target and a slice of mouse brain sample were imaged. The resolution target consists of groups with 0.5mm to 1.0mm line-width at 0.1mm interval. The size of the mouse brain sample was 10mm (Length) by 6mm (Width) after air-drying to prevent deformation during the scan. Both objects were sampled at a step size of 0.1mm in the transverse dimension, and 1° in the rotation dimension.

For comparison with flat panel detector (FPD), we also performed a CT scan on the mouse brain sample with a scintillator-based detector (1215CF-MP, Rayence). The source-side collimators were removed to directly capture each cone-beam projection. The source current was increased to 40mA to account for the low quantum efficiency of the FPD. The FPD was triggered continuously at 10 frames per second, and the first several frames (from 1 frame to 30 frames) were summed up to obtain images at different integration time settings (ranging from 0.1s to 3s), each corresponding to a different readout intensity level.

### 3.2 Reconstruction algorithm

The reconstruction solves the optimization problem in Eq. (4) for each detection scheme. Both our time-stamp photon-counting and the conventional, photon-integrating measurements are modeled by the likelihood function in Eq. (5). The measurement using the flat panel detector is modeled by Eq. (9). The reconstruction algorithm was a customized SPIRAL-TAP [16] based on the gradient of the likelihood functions. We initialized the iteration with $\mathbf{f}^{(0)} = \mathbf{A}^T \log(\lambda \mathbf{g}/\mathbf{r})$ for conventional and time-stamp photon-counting measurements; $\mathbf{f}^{(0)} = \mathbf{A}^T \log(I_0/\mathbf{y})$ for FPD measurement. The $k$-th iteration moves the solution $\mathbf{f}^{(k)}$ along the gradient direction

$$\mathbf{f}_{temp}^{(k)} = \mathbf{f}^{(k)} - \nabla l(\mathbf{f}^{(k)})/\alpha^{(k)} \qquad (12)$$

and then solves the TV de-nosing problem described in Ref. [15]

$$\mathbf{f}^{(k+1)} = \underset{\mathbf{f}'}{\operatorname{argmin}} \left\{ \left\| \mathbf{f}' - \mathbf{f}_{temp}^{(k)} \right\|^2 + \frac{2\tau}{\alpha^{(k)}} \|\mathbf{f}'\|_{TV} \right\} \qquad (13)$$

The step size $\alpha^{(k)}$ in each gradient descent step was determined according to the modified Barzilai-Borwein method described in Ref. [16], which considers the local curvature calculated from the Hessian matrix. The algorithm terminates when the relative change in the objective function between two consecutive iterations is smaller than $10^{-6}$. To prevent over-smoothing the reconstructed image, we enumerated various TV regularization parameters, $\tau$, and selected the one that yielded minimal objective function at the end of the iterations.

## 4. Results and discussion

### 4.1 Simulation results

We first performed a simulation on the conventional time-integration and our time-stamp photon-counting schemes applied to both x-ray projection and CT scenarios. The simulation Shepp-Logan phantom was a 16mm by 16mm 2D layer sampled at a voxel size of 0.2 X 0.2 X 0.2mm$^3$. PC-projection was simulated with a pixel-wise measurement. PC-CT was simulated with the experimental configurations. For conventional photon-counting scheme (Fig. 2(a, c)), each simulation instance generates binomial random numbers according to the probability in Eq. (1) to represent the photon counts $\mathbf{r}$ in a predefined period $\mathbf{g}\Delta t$. For time-stamp photon-counting scheme (Fig. 2(b, d)), the measurement, $\mathbf{g}$, is a sum of $\mathbf{r}$ geometric random numbers to represent the time intervals before the arrival of the $\mathbf{r}$-th photon.

Fig. 2(a, b) show the conventional, time-integration and time-stamp PC-CT with comparable average photon count per pencil beam. The measurement for time-integration PC-CT (Fig. 2(a)) was simulated with **g**=2048 time intervals (average 16.9 photons / beam), while time-stamp PC-CT (Fig. 2(b)) only counts the elapsed time interval of the first **r**=16 photons. Fig. 2(e) plots the log-scale normalized mean square error NMSE=$\|\hat{\mathbf{f}} - \mathbf{f_0}\|^2 / \|\mathbf{f_0}\|^2$ versus the average number of photons per pencil beam. The error bars indicate the variance of NMSE arising from 10 simulation instances of each noise model. The higher reconstruction error in time-integration photon-counting scheme is mainly attributed to the lower photon counts, thus poor SNR in the interior region of the sample. Fig. 2(c, d) compares the time-integration and time-stamp PC-projection with an average of 16 photons per pencil beam. The weak contrast features within the skull of the phantom is hardly visible at low photon counts, which agrees with the experiment result in Ref. [17]. The reconstruction NMSE versus average photon count is plotted in Fig. 2(f). In PC-radiography, the time-stamp scheme yields more uniform uncertainty among all pixels; whereas in conventional scheme, high-attenuation pixels have a larger uncertainty (and vice versa) than time-stamp scheme. Because of this, the reconstruction NMSE of time-stamp scheme is lower in low photon-counting regime ($r$<100) under the same average photon counts. Comparing Fig. 2(e) and (f), it is worth noticing that PC-CT generally outperforms PC-projection because CT absorbs more incident radiation along the beam for the same number of detected photons.

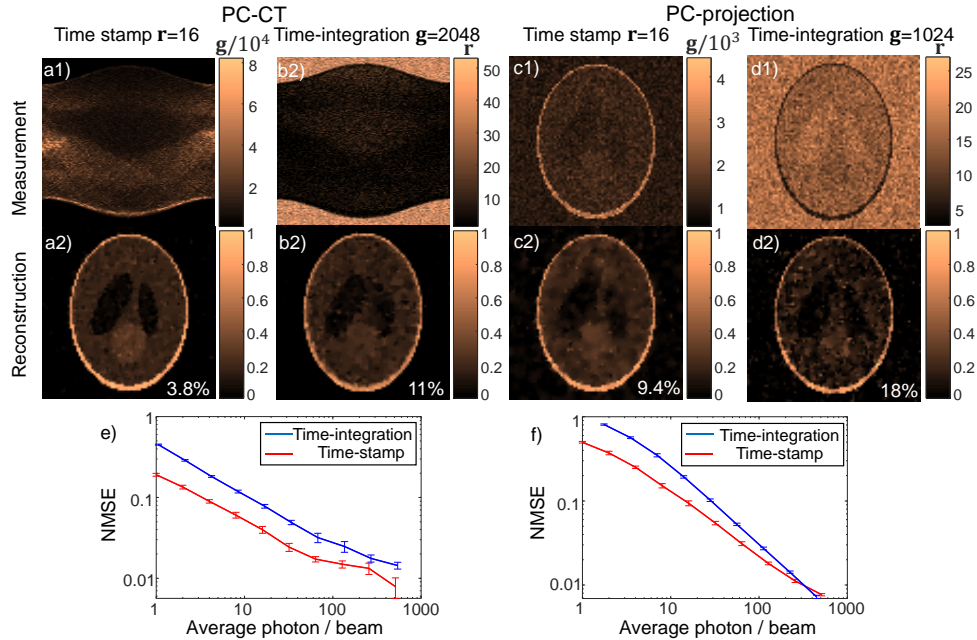

Fig. 2: PC-CT and PC-projection simulation and reconstruction of the Shepp-Logan phantom under different photon-counting schemes. (a1) time-stamp PC-CT measurement; (b1) time-integration PC-CT measurement; (c1) time-stamp PC-projection measurement; (d1) time-integration PC-projection measurement from. (a2-d2) reconstruction from (a1-d1), respectively. The numbers on each sub-figure indicate the NMSE between reconstruction and ground truth. (e, f) Log-scale plot of the reconstruction NMSE vs. the photon counts per beam for (e) PC-CT and (f) PC-radiography. All error bars indicate the variance within 10 simulation instances.

*4.2 Verification of the noise model*

The noise model was verified with an experimental measurement on the arrival time stamp of individual photons. We scanned a 2D projection (Fig. 3(a1)) covering all 9 regions of the paper pattern, and waited for the arrival of the 256$^{th}$ photon at each point. Fig. 3(a2) plots the average and variance of the time intervals, **g**, within each region in log scale. The linearity of the curve agrees with the exponential decay in the transmission as the thickness increases. From the slope in Fig. 3(a2), we estimated the transmittance, $t$=93% per paper layer. To directly observe the distribution of time intervals, we varied the number of photons, $r$, to collect at each point. Fig. 3(b1-b4) plot the histogram of **g** within region 1 at $r$=1, 2, 4 and 8. We fit a negative binomial model with one unknown, $T$, on each histogram. The red curves plot the negative binomial distributions with fitted parameter, $T$, which are 0.0127, 0.0128, 0.0129, and 0.0129 respectively in (b1)-(b4). The high consistency signifies the same photon flux exhibited on all histograms. The incident photon flux $\lambda$ was calibrated from the $T$ in region 1.

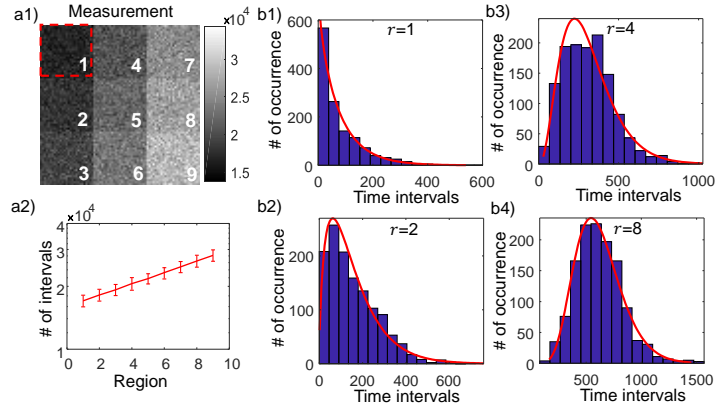

Fig. 3: Experimental observation of the photon-counting model. (a1) Number of time intervals before the arrival of 256$^{th}$ photon. (a2) The mean and variance of the time intervals in each region. (b1-b4) Histograms of the number of time intervals elapsed before $r$=1, 2, 4, and 8 photons are detected in region 1.

*4.3 Comparison with time-integration photon-counting CT scheme*

With the calibrated incident photon flux and experimentally verified noise model, a time-stamp pencil-beam PC-CT scan and reconstruction were performed on the acrylic resolution target, and compared with conventional, time-integration photon-counting scheme. Fig. 4(a1) shows the number of counts **r** in the time-integration photon-counting scheme within 1s integration time (**g**=$10^5$). The average number of photons per beam was 569. The reconstructed image (Fig. 4(b1)) was used as reference for evaluating low-photon-count images. Fig. 4(a2) shows the measurement of time-integration PC-CT with reduced integration time (0.0625s). The photon count per pencil beam was 17.8 on average. Fig. 4(a3) displays the number of elapsed time intervals **g** before the arrival of **r** =16$^{th}$ photon at each beam. The reconstructed attenuation map from time-integration and time-stamp PC-CT are shown in Fig. 4(b2, b3), respectively. The intensity profile of 0.7mm group is plotted in Fig. 4(c), which shows a visibility of 0.82 on reference image, and 0.68, 0.60 for time-integration and time-stamp PC-CT, respectively. Both time-integration and time-stamp PC-CT are capable of reconstructing small details with discernible contrast at low photon flux, because X-ray photon-counting detector eliminates the dark noise via filtration on the low-energy channels. The spatial resolution is limited by the 0.6mm spot size on the sample plane due to the beam divergence. With approximately the same average photon count (Fig. 4(b2) and (b3)), time-integration and time-stamp PC-CT have normalized mean square difference of 6.2% and 4.9% with respect to the reference image. We speculate that this slight difference is primarily attributed to the more uniform SNR on the sinogram of time-stamp PC-CT.

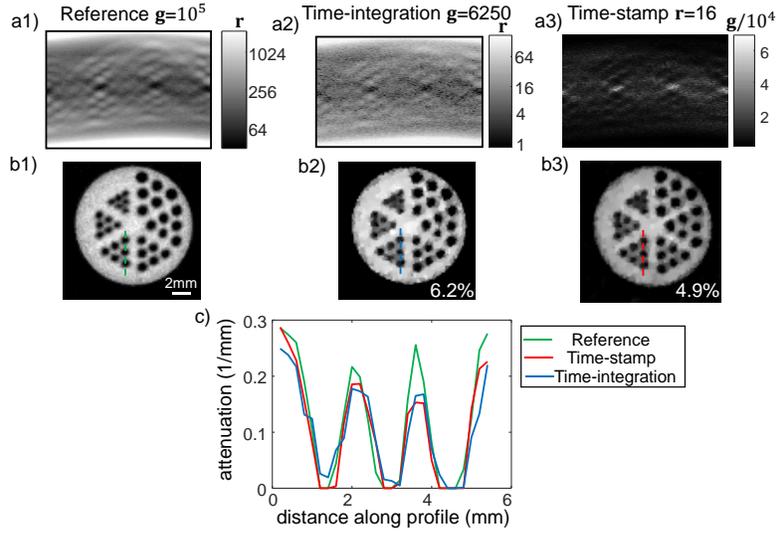

Fig. 4: Measurement (a) and reconstruction (b) of 1) reference image, 2) time-integration PC-CT scheme (17.8 photons/beam on average) and 3) time-stamp PC-CT (16 photons/beam) of a resolution target. (c) attenuation profile along the dashed line (0.7mm line-width group) in (b1-b3).

*4.4 Comparison with flat panel detector-based CT*

PC-CT has the potential in reducing the radiation dose, which is especially attractive for biomedical imaging applications. We compared the image of a mouse brain layer obtained from time-stamp PC-CT and a flat panel detector (FPD-CT, for short). Fig. 5 shows the reconstruction from FPD-CT (a, 0.5s integrating time (5 frames), 116.2 detector readout per beam on average) and time-stamp PC-CT (b, 16 photons per beam). A comparison on the absorbed radiation doses between Fig. 5(a) and (b) was performed through Monte Carlo simulation. The irradiance of the source was calculated using *XSPECT* under experimental power settings. The radiation dose of time-stamp PC-CT was calculated via an equivalent tube current modulation to simulate different integration time for each pencil beam with *ImpactMC* [18]. Fig. 5(c) shows that time-stamp PC-CT reduces the dose to ~0.6% of FPD-CT, thanks to its extremely low photon flux. It is worth noting that the dose reduction on the surface is more prominent than the center, because for CT imaging, the transmitted photon flux is intrinsically higher in the peripheral than the interior region.

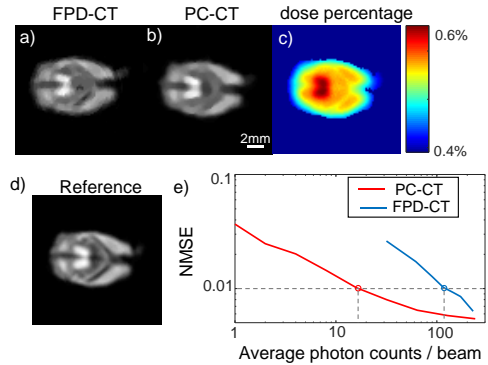

Fig. 5: Mouse brain sample imaged with (a) panel detector (FPD-CT, average 116.2 detector readout/beam) and (b) time-stamp PC-CT (16 photons/ beam). (c) Radiation dose difference between time-stamp PC-CT and FPD-CT. (d) Reference image with 1s integration time per

pencil beam (e) NMSE versus average photon counts per beam for time-stamp PC-CT and panel detector.

To further evaluate the performance between time-stamp PC-CT and FPD-CT, we acquired a complete time stamp spanning 1s integration time for the mouse brain sample using the photon-counting detector, resulting in an average photon count of 1283 photon counts per beam, and formed a reference image (Fig. 5(d)) from all the detected photons. Fig. 5(e) plots the normalized mean square difference between the reconstruction and the reference in log scale. The blue and red circles on the plot correspond to FPD-CT and time-stamp PC-CT in Fig. 5(a) and (b), respectively. For low photon counts, time-stamp PC-CT consistently performs better than conventional CT. As the photon count increases, panel detector eventually will have a comparable reconstruction error as that of time-stamp PC-CT. This is because, in high-photon flux regime, the noise model of using the panel detector and photon counting module can both be approximated by a Gaussian distribution.

## 5. Conclusion

In summary, we have demonstrated a novel x-ray photon-counting imaging scheme tailored to low photon flux scenarios. The presented method records the arrival time stamp of individual photons and reconstructs the image with a few photons per pixel, which is applicable to both x-ray projection and CT. Our photon statistics model agrees well with the actual time stamp of detected photons in the experiment. Based on this consistency, we reconstruct the PC-CT image from the arrival time stamp of the first 16 photons using a customized SPIRAL-TAP algorithm. In contrast to the conventional photon-counting scheme that records the total number of photons in a predefined integration time, our time-stamp photon-counting scheme adaptively chooses the integration time to maintain the same number of detected photons for each beam. This ensures uniform SNR across all measurements, especially for high-attenuation or interior regions on CT sinogram. The proposed few-photon method reduces the radiation dose by 2 orders of magnitude compared to CT using a panel detector. The PC-CT scheme could be extended to cone-beam. We envision the photon-counting detector array can be applied in tandem with a location addressable illumination mask, which can provide modulation to the cone beam illumination. The reduced dose opens up new opportunities in dose-sensitive biomedical or industrial non-invasive inspection applications. In addition to the presented projection and CT modalities, the photon-counting scheme can also be applied to reduce the imaging time of X-ray diffraction tomography [18], where the diffraction signal is intrinsically ~3 orders of magnitude weaker than the transmitted signal [19]. We could further exploit the energy sensitivity of X-ray photon-counting detectors to perform energy-dispersive CT or diffraction tomography for three-dimensional, *in situ* material identification.

## Acknowledgements

The authors acknowledge He Cheng from CREOL for his assistance in experimental setup.

## Funding

National Science Foundation (NSF) (DMS-1615124)